# Galactic Cosmic Rays at Mars and Venus: Temporal Variations from Hours to Decades Measured as the Background Signal of Onboard Micro-Channel Plates

## Authors

Yoshifumi Futaana[1], Manabu Shimoyama[1], Martin Wieser[1], Stefan Karlsson[1], Herman Andersson[1], Hans Nilsson[1], Xiao-Dong Wang[1], Andrey Fedorov[2], Nicolas Andre[2], Mats Holmström[1], Stas Barabash[1]

[1] Swedish Institute of Space Physics, Kiruna, Sweden

[2] IRAP, Toulouse, France

## Key points

1. The long-term measurements of the background in micro-channel plates (MCPs) in Mars and Venus orbits provide information about the local GCR intensity.

2. The modulation of the GCR intensities at Mars and Venus that are anti-correlated to the sunspot number with a time lag of 9 months for the cycle 24.

3. There are short-term modulations in the GCR intensity due to the blockage with planetary bodies and their atmospheres

## Abstract

A Micro-Channel Plate (MCP) is a widely used component for counting particles in space. Using the background counts of MCPs on Mars Express and Venus Express orbiters operated over 17 years and 8 years, respectively, we investigate the galactic cosmic ray (GCR) characteristics in the inner solar system. The MCP background counts at Mars and Venus on a solar cycle time scale exhibit clear anti-correlation to the sunspot number. We conclude that the measured MCP background contain the GCR information. The GCR characteristics measured using the MCP background at Mars show features that are consistent with the ground-based measurement in solar cycle 24. The time lag between the sunspot number and the MCP background at Mars is found ~9 months. The shorter-term background data recorded along the orbits (with a time scale of several hours) also show evident depletion of the background counts due to the absorption of the GCR particles by the planets. Thanks to the visible planetary size change along an orbit, the GCR contribution to the MCP background can be separated from the internal contribution due to the β-decay. Our statistical analysis of the GCR absorption signatures at Mars implies that the effective absorption size of Mars for the GCR particles have a >100 km larger radius than the solid Martian body.

## 1  Introduction

High energy particles exist ubiquitously in the universe. One of the most notable examples is galactic cosmic rays (GCRs). Early in the 20th century, research on GCR particles progressed our understandings of high-energy physics thanks to their typical (peak) energy of ~GeV. Later, GCR research moved on to understanding the early history of the universe. Currently, GCR study is used for heliospheric science through investigating the modulation of GCR intensity with geomagnetic disturbances and solar activity (Forbush 1938; Van Allen 1993; Potgieter 2013). Another example of high-energy particles is solar energetic particles (SEPs). SEPs are produced by shocks in front of coronal mass ejections (CMEs) related to flare eruptions or coronal interaction regions (CIR). The energy of SEPs is typically in the MeV range (Reames 1999; Ryan et al. 2000).

These high-energy particles are a threat to human activities in space. Robotic missions suffer electronics failures from high-energy particles. Humans in space also suffer from these particles via damage to deoxyribonucleic acid (DNA). Therefore, monitoring high-energy particles such as GCRs and SEPs is critical for space exploration, and the characterization of these particles has been of interest to the space weather community. Since deep-space cruising such as travel to the Moon or Mars is foreseeable in the future, the demand of understanding the high-energy particle radiation environment in interplanetary space and near extra-terrestrial planets is increasing.

In this study, we investigate the temporal variations of GCR intensity using the micro-channel plane (MCP) backgrounds operating near Mars and Venus over 17 years and 8 years. It is well-known that high-energy particles provide an MCP background count. For example, SEP particles produce this type of background. We already reported that the MCP background represents the SEP intensity at Mars and Venus (Futaana et al. 2008; Ramstad et al. 2018). With the same principle, GCRs are recorded as the MCP background counts. This paper focuses on the GCR intensity near planetary bodies (Venus and Mars) and the temporal variations with different time scales, from hours to decades.

## 2 Micro-Channel Plate background

An MCP is a widely used component for counting photons and particles (Carlson and McFadden 1998). MCPs have advantages since they have wide effective areas and high precision for determining the position where incoming particles hit on the plate. However, MCPs are also known to produce background counts (Siegmund et al. 1988). In this section, we briefly overview the characteristics of MCP background sources.

High-energy particles are a notable source of the MCP background. In space, the main contribution is from SEPs and GCRs. Electrons from the Jovian magnetosphere add a minor contribution to the background. Radiation sources for space vehicles (e.g., radioisotope thermoelectric generators (RTGs) or radioisotope heater units (RHUs)) also produce the MCP background. While MCPs are usually located deep inside space instrumentation, high-energy particles can penetrate the instrument wall and directly access the MCP. Some particles stochastically collide with atoms inside the MCP lattice to produce a background signal. A minor portion of the MCP background signal can be caused by secondary species (including photons). These are produced by the impact of primary high-energy particles on the structures of the instrument and spacecraft.

SEPs are the electrons and protons accelerated by shock waves formed by solar flares, interplanetary coronal mass ejections (ICME) and corotation interacting region (CIR). These particles can typically reach MeV-level-energies (e.g., Reames 1999; Ryan et al. 2000). In most cases, SEP fluxes are abruptly increased by several orders of magnitude when an ICME or CIR arrives at the spacecraft location. The time series of the MCP background is quite similar to the general trend of the SEP intensity. With consideration of these characteristics, we used the background counts of MCPs for space weather monitoring purposes through the heliosphere (Futaana et al. 2008; Ramstad et al. 2018; Palmerio et al. 2021).

GCRs are another population of high-energy particles in space. GCRs mostly come from outside the solar system, mainly in the Milky Way. They are thought to be created by supernova explosions (Simpson 1983). In addition, a small portion of GCRs may have been produced in other galaxies (Aab et al. 2017). GCRs usually have higher energy than SEPs, peaking at several hundreds of MeV to several GeV. They are mainly proton and alpha particles, with minor constituents of high mass species.

It is well-known that a clear solar cycle modulation (anti-correlation) exists between the GCR intensity and the solar cycle (Van Allen 2000; Potgieter 2013). This modulation is caused by the different states of turbulence in the solar system for the different solar activities, resulting in the GCR transport disturbances. Geomagnetic disturbances due to magnetic storms also modulate the GCR intensity at the ground (known as Forbush effects; Forbush 1938). Recently, Forbush effects have been recorded at the Martian surface driven by CME passages, and these Martian Forbush effects were highly correlated with the events on Earth (Papaioannou et al. 2019).

High-energy electrons leaking from the Jovian magnetosphere can also contribute to the MCP background in the inner solar system (Krimigis et al. 1975). The intensity of the Jovian energetic electrons exhibits ~13-month modulation at Earth because the Earth and Jupiter are magnetically connected every ~13 months (Chenette et al. 1977).

Regardless of the source of the high-energy particles, deriving the absolute intensity or the shape of the energy spectra from MCP background data is generally not straightforward. Assuming an energy spectrum of high-energy particles (the differential flux, with the units in cm$^{-2}$ s$^{-1}$ sr$^{-1}$ eV$^{-1}$), $\mathbf{j}(E, \mathbf{\Omega})$, the background counts (C) are expressed as

$$C = \iint \mathbf{j}(E, \mathbf{\Omega}) \, P(E, \mathbf{\Omega}) \, dE \, d\mathbf{\Omega} \quad \ldots (1)$$

where $P(E, \mathbf{\Omega})$ is the probability of producing the counts when a single high-energy particle (with energy E coming from the direction, $\mathbf{\Omega}$) hits the MCP. Because we can only determine the integrated value C from the background measurement in space, we should deconvolve the differential flux. This is a typical inversion problem in the underdetermined system. A further complication comes from the fact that we have no information about $P(E, \mathbf{\Omega})$. Theoretically, $P(E, \mathbf{\Omega})$ is non-linear to the energy. It also depends on the materials surrounding the MCPs. Moreover, the shapes of the instrument and the spacecraft matter.

The β-decay of radioactive elements in MCPs is another strong source to produce the background. With this process, MCPs produce signals without interacting with particles in space. The internal sources are from radioactive elements inside the MCP glass. When the radioactive element decays, electrons are emitted inside the MCP, which causes electron avalanches in the MCP, producing the background.

The most significant contribution comes from $^{40}$K. The half-life is ~10$^9$ years, so the contribution is constant over the temporal scale of 10 years in this study. Previous studies (e.g., Siegmund et al. 1988) reported that the typical MCP background event rate was ~0.1 cm$^{-2}$ s$^{-1}$.

There are several other processes that can create the background. Siegmund et al. 1988 explained that outgas from an MCP can produce background counts. A hotspot produced by a dust particle on an MCP could also contribute to the background. However, these contributions are presumably minor for space instrumentation because the sensors are baked and cleaned adequately before launch. After launch, instruments are operated in a high vacuum environment.

UV photons are an additional possible source of the background. MCPs are usually located inside instrument boxes, and photons entering their apertures should experience several collisions with properly coated walls before reaching the MCPs. Thus, the UV contribution is less important in general.

## 3    Instruments and data

We use the datasets obtained from the European Space Agency's (ESA's) planetary missions, Mars Express and Venus Express. Mars Express was launched in 2003 to explore the red planet (Chicarro et al. 2004). The Mars Express spacecraft was inserted into a Martian orbit (with a pericenter of ~250 km and an apocenter of ~10000 km, both measured from the surface of Mars). The mission is still in operation (as of 2022) orbiting around Mars, and we expect further continuous operation. Measurements over 17 years around Mars have been achieved, and a unique dataset that is useful for understanding our neighboring planet and the space environment near the planet is available.

Venus Express was launched in 2006 to explore our twin planet, Venus (Svedhem et al. 2007). Venus Express was inserted into a highly elliptical polar orbit with a nominal pericenter altitude of ~200 km. Eventually, the pericenter was lowered to 135 km (Muller-Wodarg et al. 2016). Venus Express was operated until the end of 2014, when the spacecraft consumed the entirety of its fuel to maintain the orbit. We have more than 8 years of data for the orbit around Venus.

Among the many scientific instruments on board Mars Express and Venus Express, the space plasma instruments, Analyser of Space Plasma and EneRgetic Atoms (ASPERA-3 and ASPERA-4), were carried to Mars and Venus (Barabash et al. 2006; Barabash et al. 2007). These sensors were based on a common design but tailored to each environment. These instruments have been used for scientific investigations to reveal the solar wind interaction with the non-magnetized planetary atmospheres. A specific focus has involved quantifying the atmospheric escape in the form of charged particles (Futaana et al. 2017). In addition to the charged particle measurements, the first-ever remote sensing of the plasma environment was conducted by measuring energetic neutral atoms (ENAs) (Futaana et al. 2011).

In this study, we use the Ion Mass Analyser (IMA) as a part of the ASPERAs. The maximum energy that can be resolved by an IMA is ~36 keV/q. No high-energy plasma instruments data are available because these missions were not designed to investigate high-energy plasma physics. However, we can use the background counts of the IMAs to monitor high-energy particles. If we assume that 2 mm of aluminum equivalent shielding as a rough estimation, ~25 MeV protons can penetrate through (NIST 2017), which provides the first-order estimation of the minimum energy of protons that produce the IMA background. Inter-comparison with high-energy particle data onboard the Mars Atmosphere Volatile EvolutioN (MAVEN) spacecraft provided the threshold of the contributing particle energy experimentally. The main contributions of the particles are for IMA MCP background electrons with >1 MeV, and protons with >~20 MeV (Ramstad et al. 2018).

Figure 1a shows the cross-section view of an IMA. The IMA sensor is composed of the following four main functionalities.

(1) An electrostatic deflection system to select the elevation angle (the angle with respect to the aperture plane of the top-hat electrostatic analyzer).

(2) An electrostatic analyzer (ESA) for energy filtering.

(3) A velocity analyzer based on a permanent magnet system.

(4) An MCP detector (Figure 1b) with a position-sensitive anode system.

By setting specific voltages between the upper and lower electrodes of the deflection system, only ions coming from a particular direction can get through the deflection system. By sweeping the voltages, we can obtain the incoming direction information in the range of ±45˚. For ESA, with a specific voltage setting between the inner and outer spherical electrodes, only the ions with a proper energy range can get through. By sweeping the voltage between electrodes, IMA can provide the energy spectra of ions, with a maximum energy range from 10 eV/q to 36 keV/q. The range can be programmable by a pre-defined table. At the end of the electrostatic analyzer, the ions are accelerated (post-acceleration) by a specific negative bias. Inside the velocity analyzer part, the magnetic field in the azimuth direction is produced by permanent magnets, and the ions are bent according to their velocity. Because of the confined incoming ion energy, the velocities of the ions differ according to the masses ($v=(2E/m)^{1/2}$, with v being the speed, E being the energy, and m being the mass of ions). Thus, the trajectory of the ions in the velocity analyzer depends on the mass. Finally, the ions impact the MCP (two-stacked plates with a 100 mm diameter), which is biased by –2800 V (Barabash et al. 2006). The impact point of an ion depends on its mass. An MCP produces ~$10^5$–$10^6$ electrons that are fed to a position detection system behind the MCP. The position detection anode system provides the impact position with 32 radial (rings) and 16 azimuthal (sector) resolutions using six preamplifiers and an FPGA-based coincidence algorithm. We can obtain the histogram of counts with 32 (mass) times 16 (azimuthal angle) bins every 125 ms. The energy scan (ESA, 2) is performed with 96 steps (12 s cadency), followed by the elevation angle scan (deflector, 1) with 16 steps. Onboard processing creates a four-dimensional (4-D) histogram of the counts (mass, energy, azimuth angle, and elevation angle) every 192 s.

The time series of a 4-D histogram is the starting point of this study. Then we reduce the dimension of the histogram to 2-D (the energy-mass histogram) by collapsing the information about the directions. To observe the statistical features of the IMA performance, we show Figure 2 as the example of the energy-mass histogram accumulated over one month. The number of mass bins is 32, and that for of the energy is 96. The mass bin index corresponds to the radial position of the MCP (ring). The energy index number corresponds to the physical energy per charge (indicated on the right axis). Depending on the actual mass-per-charge and the actual energy of the incoming ions, the foreground ions appear specific areas in the energy-mass plot, as represented by the white curves in this figure.

The background counts are visible as a spread of constant counts over the whole bin. On top of the background, the foreground signal is clearly seen as clusters of counts. For example, the solar wind protons, with energies of 0.5–2 keV/q, are seen along the M/q=1 curve, on the mass ring number >25, in the Mars Express data (Figure 2a). A weaker signal at M/q=2 (solar wind alpha particles) is also seen at 1–4 keV/q. The ionospheric ions from Mars are found at the lowest energy with mass bins 8-15. The visible depletions at specific mass ring numbers (0, 4, 11, and 22) are artificial: we disabled those mass bins on board due to substantial interferences in the position detection system (frontend electronics). Electronics interference also causes a moderate difference in the efficiency, e.g., the background signal for the mass bins 16, 20, 24, and 26 is higher than that for the neighboring bins. The improvement of the frontend electronics for the Venus Express IMA provided better performance in terms of the detection efficiency dependence on the mass channel (Figure 2b). Of course, the foreground ions (solar wind proton, alpha, and Venusian ionospheric ions) can easily be identified in the Venus Express data as well.

Since the M/q of any ion species is limited to ≥1, no foreground ions can reach the upper right corner areas in the energy-mass plots (Figure 2). We took the region as the background channel. The optimal background channel for the Mars Express IMA consists of the mass bins 27–30 and the energy steps 1–11. For the Venus Express IMA, the mass bins 25–30 and the energy steps 1–12 are used (Capalbo 2010). To derive the background counts, we take the sum of the counts inside the background channel. The summed background count (every 192 s) is then divided by the number of bakcground bins (44 for Mars Express and 84 for Venus Express), producing the background count raw data ($BG_{MEXRAW}$ and $BG_{VEXRAW}$).

These background counts are highly related to the instrument design, and it is not straightforward to convert them to the physical quantities. Due to the lack of knowledge and calibration data, the effective areas of the MCP are only slightly constrained. Our simple estimation provides ~18 cm² and ~27 cm² as the effective areas corresponding to the background channel for Mars Express and Venus Express, respectively. The effective duration of the measurements are 22 s for Mars Express and 28 s for Venus Express during the cadency of 192 s. Overall, the rough count rate of the background in a physical unit may be calculated from the raw background count values.

$BG_{MEX}$ [cm$^{-2}$ s$^{-1}$] ~ 0.11 x $BG_{MEXRAW}$     ..... (2)

$BG_{VEX}$ [cm$^{-2}$ s$^{-1}$] ~ 0.11 x $BG_{VEXRAW}$     ..... (3)

However, we keep the raw background count values for this study (unless otherwise noted) to avoid readers receiving the impression that the numbers shown have been well-calibrated.

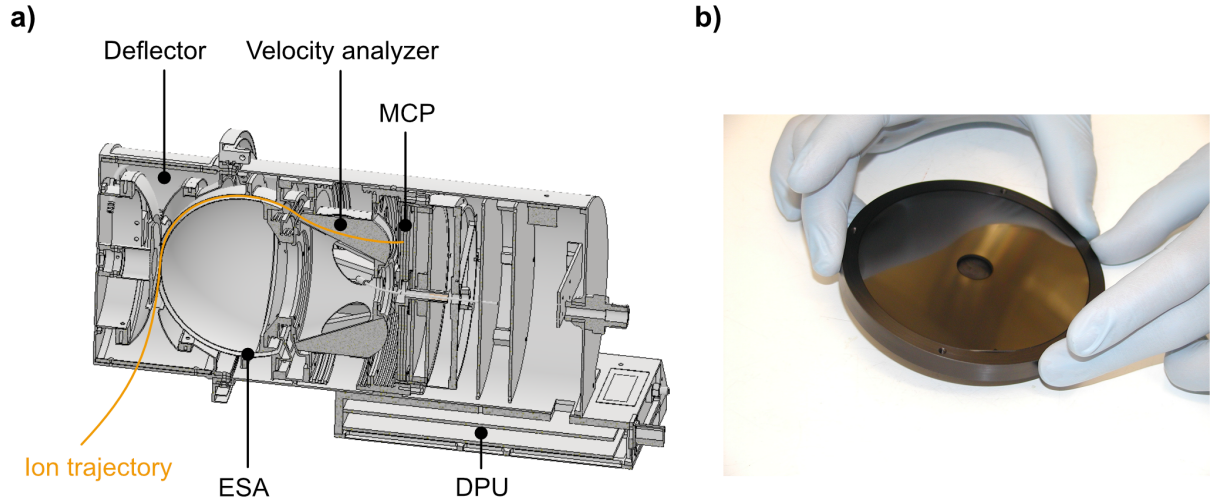

Figure 1: a) A cross-section cut of the Ion Mass Analyser (IMA) instrument with four main sections: Electrostatic deflector, electrostatic analyzer (ESA), magnets, and microchannel plate (MCP). The Digital Processing Unit (DPU) is the processing unit of IMA (Barabash et al. 2006). b) MCP used for Mars Express IMA with a diameter of 10 cm.

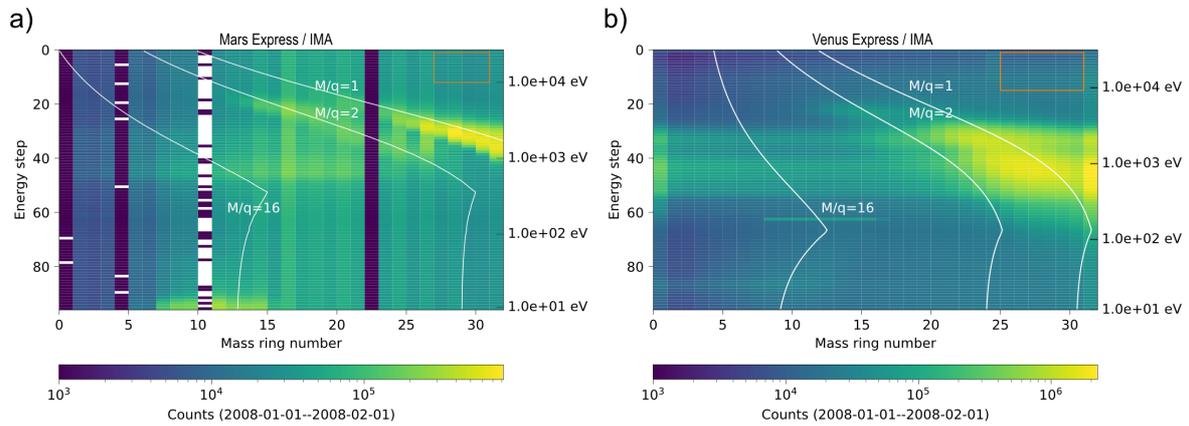

Figure 2. a) The Mars Express and b) Venus Express IMA counts accumulated over 1 month (January 2008) represented in the energy-mass matrix. The energy step contains 96 bins. The corresponding foreground ion energy is shown on the right. The horizontal axis is the mass ring number. The white curves represent the foreground ion's mass-per-charge and the energy (proton, alpha, oxygen ions, and other positive ions species). The mass bins 0, 4, 10, and 23 for Mars Express were disabled due to substantial interferences. The background counts are spread over all the energy and mass bins. The orange boxes indicate the background channels.

## 4    Decadal variations

This section describes the long-term trends of the IMA background over the 11-year solar activity cycle. Figure 3 shows the time series of the background data. Every three months, the median values of the background counts are shown. The typical number of IMA scans is ~10000 within each three-month window. In addition to the median, the 5%, 25%, 75%, and 95% percentiles are also indicated in the figure. By taking the median, we can minimize the influence from SEP events because the intense SEP flux only continues for approximately a week.

The Mars Express IMA background counts are 1–3 per scan, corresponding to ~0.1–0.3 $cm^{-2}$ $s^{-1}$ (Equation 2). The Venus Express IMA background counts are 6–8 per scan, corresponding to ~0.7–0.9 $cm^{-2}$ $s^{-1}$ (Equation 3). The nominal expected background count rates of 0.1 $cm^{-2}$ $s^{-1}$ due to the β-decay for typical MCPs and the GCR contribution at the ground level are typically 0.1–0.5 $cm^{-2}$ $s^{-1}$ (Siegmund et al. 1988). All these quantities are within the range of the IMA background count rates.

A distinct signature found in Figure 3 is a clear trend of anti-correlation between the sunspot number (Figure 3c) and the Mars Express IMA background (Figure 3b). The anti-correlation is consistent with the GCR characteristics at Earth, the intensity of which is known to anti-correlates to the solar activity (e.g., Van Allen 1993; Van Allen 2000). We can conclude the substantial contribution of GCR intensity to the MCP background at Mars.

In the Venus Express IMA background data (Figure 3a), the anti-correlation to the sunspot number is less visible. In the first year (2006), the background shows a decreasing trend. This feature is probably due to the outgassing influence on the MCP background. Space instruments continuously outgas in the extremely high vacuum environment after launch (Schlappi et al. 2010) while they were baked out on ground. After 2007, a monotonic increase of the background is seen as the sunspot number decreases. This increase is consistent with the Mars Express IMA background increase from 2007 to 2010. A distinct drop in the Venus Express IMA background after 2010 is due to the change of the onboard processing. We introduced an onboard background subtraction scheme to increase the primary science outcome. In fact, we already knew in 2006 that relatively high background counts had increased the load in the onboard IMA processing system, resulting in packet losses of the data. To avoid data losses, the nominal 4-D matrix was re-binned by reducing the number of elevation bins from 16 to 8. The drawback was the low angular resolution in the data. In 2010, a background subtraction scheme was introduced instead of re-binning. The counts in all the bins were decremented by 1 onboard. Therefore, the bins with the raw count 1 were reported as 0 counts, and the load on the system was significantly mitigated. However, the background counts were reduced considerably, as shown in Figure 3a. We further show in Figure 4 that the reduced background count values are still anti-correlated to the sunspot number. This confirms that the GCR intensity at the Venus location is also modulated by the solar activity, but a further systematic long-term analysis is challenging.

It would be an exciting discussion to determine how much the GCR flux differs between Mars (1.5AU) and Venus (0.72AU). Honig et al. 2019 concluded that there was a 3 % / AU gradient in the GCR intensity between 1 AU and 4.5 AU. If we apply the decrease rate, only a couple of percent of decrease is expected between Venus and Mars. However, comparing the absolute count rates of the Mars Express and Venus Express IMA background is complex because the detailed implementation was not entirely identical. For example, the frontend electronics for these sensors were optimized to perform the foreground (space plasma) measurements in different environments. The thresholds are different, and the sensor temperature during operation also differ. In addition, since the background measurements were not an objective of the mission and these sensors, we did not conduct any ground calibrations for high-energy particles. Therefore, the response of each MCP to the high-energy particles is not known. Furthermore, the housings of the MCP (shape of the spacecraft and the IMA sensor body) differ. In summary, we cannot quantitatively discuss the difference between the Venus Express (~0.7–0.9 $cm^{-2}$ $s^{-1}$) and Mars Express (~0.1–0.3 $cm^{-2}$ $s^{-1}$) IMA background.

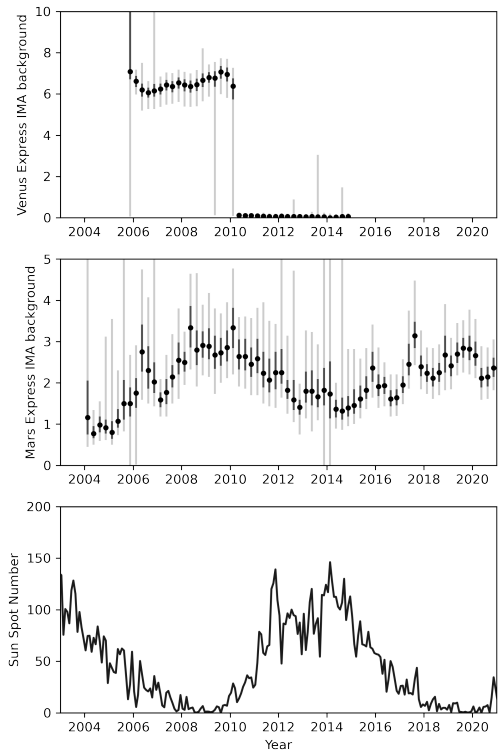

Figure 3: The time series of the background counts of the (a) Venus Express IMA and (b) Mars Express IMA. The dots represent the median values derived from all available data during the three-month interval regardless of the spacecraft position and the attitude. The error bars are for the 25% and 75% percentiles (black) and 5% and 95% percentiles (gray). For the Venus Express IMA, an onboard background subtraction scheme was introduced in 2010 so that the background is extraordinarily small after 2010. (c) Monthly sunspot number provided by the World Data Center SILSO, Royal Observatory of Belgium, Brussels.

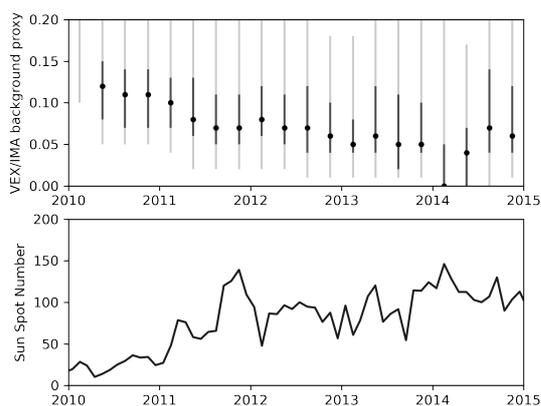

Figure 4: Time series of the Venus Express IMA background after 2010. Due to the onboard background subtraction, the background count value is deficient. However, one can still see the anti-correlation to the sunspot number, indicating that a contribution of the GCR is visible in the reduced background data at Venus.

It is well known that there is an anti-correlation between the sunspot number and the GCR intensity. This anti-correlation is due to the diffusion of GCR particles due to the turbulence in the heliosphere.

Interestingly, the signature of the GCR intensity differs between the odd-numbered solar cycles (e.g., Cycles 21 and 23) and the even-numbered cycles (Cycles 20 and 22) (e.g., Van Allen 2000; Belov 2000). This difference is explained by the large-scale solar magnetic field polarity difference. However, an exception occurred recently in the cycle 24. Cycle 24 showed a similar signature to the odd-numbered cycles (Ross and Chaplin 2019).

Here we compare the Mars Express IMA background and a dataset for the neutron monitor observed at the Thule station (76.5° N, 68.7° W). Neutron monitors have been widely used for GCR monitoring because a GCR particle produces a neutron shower in the middle atmosphere. Figure 5 shows the hysteresis of the Mars Express background counts (1-year median) and the neutron intensity measured at the Thule station compared with the sunspot number. We can see similar and consistent hysteresis shapes.

Another interesting feature is a possible time lag of the Mars Express background against the sunspot number (Figure 3b and c). The recovery of the GCR intensity appears to be delayed after the peak of solar activity. In many previous studies, time lags of ~10–12 months (e.g., Kane 2014) have been known to exist for odd-solar cycles. However, even-numbered cycles generally do not show time lags. The only exception ever identified is for cycle 24, where a ~4-month lag was reported using ground-based neutron monitors (e.g., Tomassetti et al. 2017; Ross and Chaplin 2019). Recently, using the engineering parameter of the error detection and correction (EDAC) counters of Mars Express, Knutsen et al. 2021 derived the 5.5-month time lag between the GCR intensity and the sunspot number for the years 2005–2020 at Mars. The exceptional feature in cycle 24 is attributed to the long solar minimum between cycles 23 and 24 or the relatively low solar activity of cycle 24.

Figures 6 and 7 show the correlation coefficients between the Mars Express IMA background (three-month median) and the time-shifted sunspot number (three-month average). Figure 6 displays the best correlation for the nine months of lag with a correlation coefficient of –0.86. Figure 7 shows the data with the nine-month time lag. The lag is slightly longer than the previously reported 4-month lag from the ground-based measurements and the 5.5-month lag from the Mars Express EDAC counters. In fact, the 9-month lag is more consistent with that identified for odd-numbered cycles (Ross and Chaplin 2019). The effective energy of GCR particles to produce the background may be different from ground-based observations or the EDAC counter because the time lag depends on the GCR particle energy.

Nevertheless, the correlation is high in general (<-0.7) (Figure 6). Because of the lack of the Mars Express IMA background characterization, our result may not dispute the previous results. We emphasize that our results support the existence of time lag for cycle 24, indicating that the cycle 24 was exceptional among other even-number cycles, as shown in previous studies.

In summary, the Mars Express IMA background data show clear anti-correlation to the solar activity, with a time lag of nine months. We can conclude that the GCR intensity signatures are common at Earth (1 AU) and Mars (1.5 AU). We carried out a similar investigation using the Venus Express IMA background data. However, the onboard background reduction scheme (Figure 4) made the long-term analysis significantly complicated. Because of the change of operation at the time of the peak solar activity, the time lag analyses, unfortunately, did not result in meaningful results.

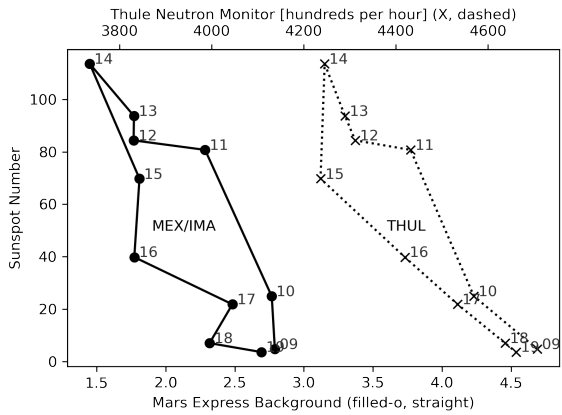

Figure 5: The hysteresis plot for the Mars Express IMA background at Mars and the Thule neutron monitor on the ground (76.5˚N, 68.7˚W). The data for the Thule neutron monitor were provided by the University of Delaware Department of Physics and Astronomy and the Bartol Research Institute.

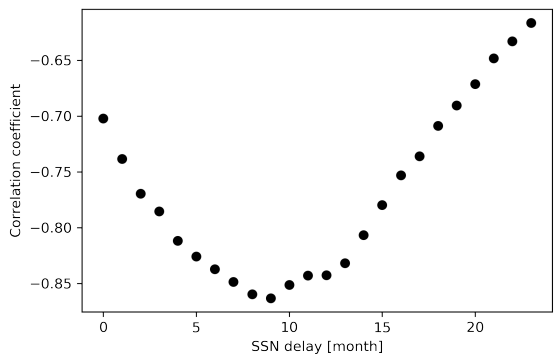

Figure 6: The results of the correlation study between the sunspot number and the Mars Express background for solar cycle 24 (2009-2019). The best agreement (anti-correlation) is found at a nine-month lag, while the correlation coefficient is good (<-0.7) in general.

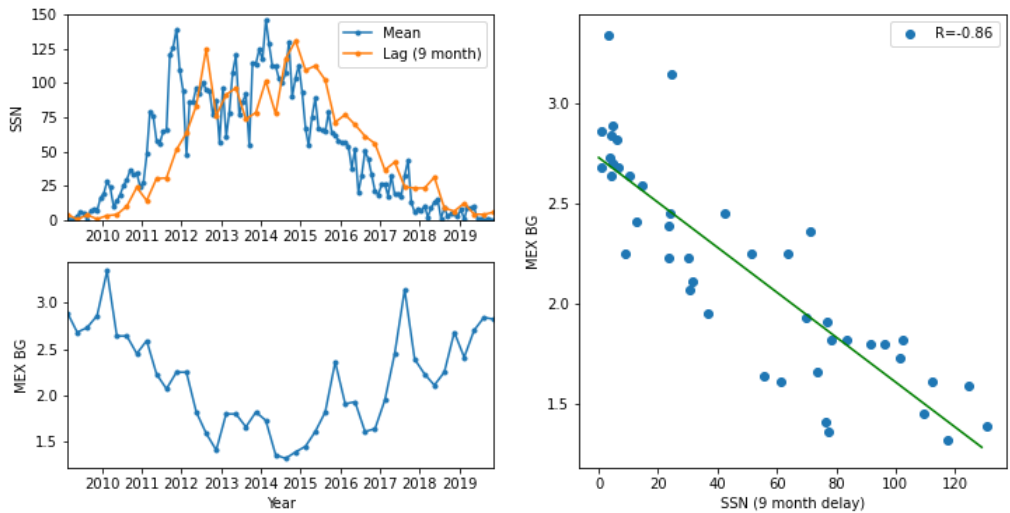

Figure 7: (Top left) The time series of the sunspot numbers (monthly average, blue) and that with the 3-month average with the 9-month time lag. (Bottom left) The time series of the measured Mars Express background. (Right) The scatter plot between the 9-month delayed sunspot number and the Mars Express background. The linear regression line is drawn by the green line with a correlation coefficient of –0.86.

## 5   Orbital modulation

In contrast to the previous section where we discussed the decadal modulation of the MCP background due to solar activity, this section focuses the shorter-time modulation with several hours. We can expect a change of the GCR-induced count rate along an orbit because the closer the spacecraft is to the planets, the more GCR particles are absorbed by the planet. On the other hand, the internal counts driven by the decay of radioisotopes are constant as the half-time is significantly long (>$10^9$ years). Thus, we can separate the influence of the external GCR source and the internal β-decay source from the time series of the background data.

We develop a simple model to represent the background counts. The model is applied to the time series of the background data along the trajectory of the spacecraft over one orbit near the pericenter. Our model assumptions are as follows:

1. The internal background is constant over the one pass of the spacecraft (several hours).

2. The external source (GCR particles) is isotropic, and the characteristics (e.g., energy, flux) do not change over one pass.

3. The external source coming from the direction of the respective planetary body is entirely blocked.

Because of the long half-time of $^{40}K$, the first assumption would be justified, while the electronics in the detection system may contribute to modifying the constant count rate possibly due to the temperature drift of the threshold. In our simple model for this study, we cannot consider these types of non-linear effects.

Using the above assumptions, we first formulate the GCR-induced counts. From the isotropic GCR differential flux (without planetary blockage) denoted as j [$cm^{-2}$ $s^{-1}$ $sr^{-1}$] (already integrated over the energy), the background counts detected by a planer MCP, denoted dC, contributed from a solid angle (d$\Omega$) is calculated by

$$dC = j\ S\ (\mathbf{n} \cdot d\mathbf{\Omega})\ r(\theta, \varphi) = j\ S\ d\Omega\ \cos\theta\ \varepsilon(\theta, \varphi) \quad \ldots\ (4)$$

Here, S is the area of the MCP (which is constant in this case), $\mathbf{n}$ is the normal vector of the MCP, and ε is the response of MCP to the GCR particles that comes from a specific direction. The angle θ is the polar angle (between the normal vector $\mathbf{n}$ and the direction of interest), and φ is the azimuthal angle. We assume the system is axisymmetric, i.e, there are no dependences on φ. The actual formulation of ε is unknown, but a simple model is used in this study: the 1/cos θ dependence. The motivation of this assumption is as follows. The high-energy GCR particle (~GeV) can penetrate the MCP. Therefore, the probability of a collision between a GCR particle and an atom in the MCP producing electron avalanches is proportional to the pass length of the GCR particle through MCP. The path length is the 1/cos θ dependence. This simple formulation may only represent an ideal case, but in fact, a similar trend in the angular response of MCP was reported against the X-ray photon (5-20 keV) (e.g., Kondoh et al. 1988).

Using the 1/cos θ dependence, the MCP count is simplified to

$$dC = j\ S\ R\ d\mathbf{\Omega} \quad \ldots\ (5)$$

where R is the response for the particle coming from normal (θ=0), which is constant. Equation (5) clearly shows that the differential counts of the GCR-driven background are proportional to the solid angle of interest (d$\Omega$). Therefore, if a planet blocks the GCR particles, the reduction in the background counts is proportional to the solid angle of the planet seen from the MCP. The blockage solid angle $\Omega_p$ is pure-geometrically 2π(1-cos α) with α=arcsin ($R_p/r_{sc}$). Here, $R_p$ is the planetary radius and $r_{sc}$ is the spacecraft distance from the planetary center. Taking the reduction by a planetary blockage into account, the GCR contributing counts are:

$$C_{GCR}(r_{sc}) = B * (4\pi - \Omega_p) / 4\pi \quad \ldots\ldots (6)$$

The parameter B represents the total counts due to GCR expected without the planetary blockage. Because the spacecraft distance $r_{sc}$ changes over time, the contribution from the GCR ($C_{GCR}$) also changes. The total background counts are thus simply:

$$C_{bg}(t) = A + B(1+\cos\alpha)/2 \quad \ldots\ldots (7)$$

$$\alpha = \arcsin(R_p/r_{sc}(t)) \quad \ldots\ldots (8)$$

Here A is the background count from β-decay that is assumed constant.

By fitting the time series of the measured total background counts against the model (Equation 7), we can separate the internal (A) and external (B) sources from a single orbit. The background counts and each contribution should always be a positive value. We thus constrain both the parameters to be always positive (i.e., A≥0 and B≥0) for the fitting.

Figure 8 shows the typical background counts overlaid with the fitting results using the above simple model ($R_M$=3396 km and $R_V$=6051 km). A clear trend in the background intensity is found that is dependent on the spacecraft's distance to Mars and Venus. The lowest background counts are seen in Figure 8a at the pericenter of ~250 km, when the apparent size of Mars is maximum. For Venus, precisely the same signature is found (Figure 8b). The depletions in the near peri-center indicate that the shadowing effect by planets is also in operation. The best-fit parameter pair (A, B) indicates the relative contribution between the constant internal and time-varying external sources following the Equation (7).

We can obtain one best parameter pair (A, B) from each pericenter pass. Further statistical analyses are possible. Out of the 21550 orbits of the Mars Express from 2003 until January 18, 2021, we use the orbit when the continuous measurements are available around the pericenter. The criterion is that the number of the background data point is more than 50 (~160 min) within three hours around the pericenter time. We have 7371 orbits that satisfy this condition.

Figure 9a shows the statistical distribution of the internal (A) and external (B) sources from the Mars Express IMA background. For the fitting, the constraints A≥0 and B≥0 are applied to restrict the fitting because both components cannot negatively impact the background intensity. Surprisingly, many of the best-fit parameters are found at A=0 (Figure 9a). This result implies that the time-variation represented by the external GCR should influence more significantly than the simple geometric model.

Therefore, we perform further examination by increasing the planetary radius (considered an obstacle sphere to GCRs) by 100 km, 200 km, and 300 km (Figure 9b-d). The cluster of the best (A, B) pair is found at higher A and smaller B, and the amount of data at A=0 is significantly less. While the quantitative assessment is not straightforward, these results imply that the effective blocking radius has to be >100 km larger than the Martian radius.

It may not be surprising that the effective blockage size tends to be larger than the actual Martian body. While Mars has no magnetic field, it does have an atmosphere. The atmospheric effect can reduce the GCR flux toward the ground (Townsend et al. 2011; Ratliff et al. 2017). Further comparisons with the Radiation Assessment Device (RAD) measurement on the Curiosity rover (Hassler et al. 2012), as well as the MAVEN SEP instrument (Larson et al. 2015), are a natural development of this study.

The same exercise is performed for Venus (Figure 10). Most of the best-fit (A, B) pairs are seen already in the proper range without increasing the size of the effective blockage. Increasing the effective blocking radius does not provide sensitive results. We thus conclude that it is not possible to investigate the effective size of the atmospheric shadowing for Venus. The question then arises of why it is different from Mars. One possible interpretation is that Venus has a larger radius than Mars, although the atmospheric scale height is similar. Thus, the effect of increasing the radius is relatively minor for Venus, and the background data could be insensitive to the increase of the blockage radius.

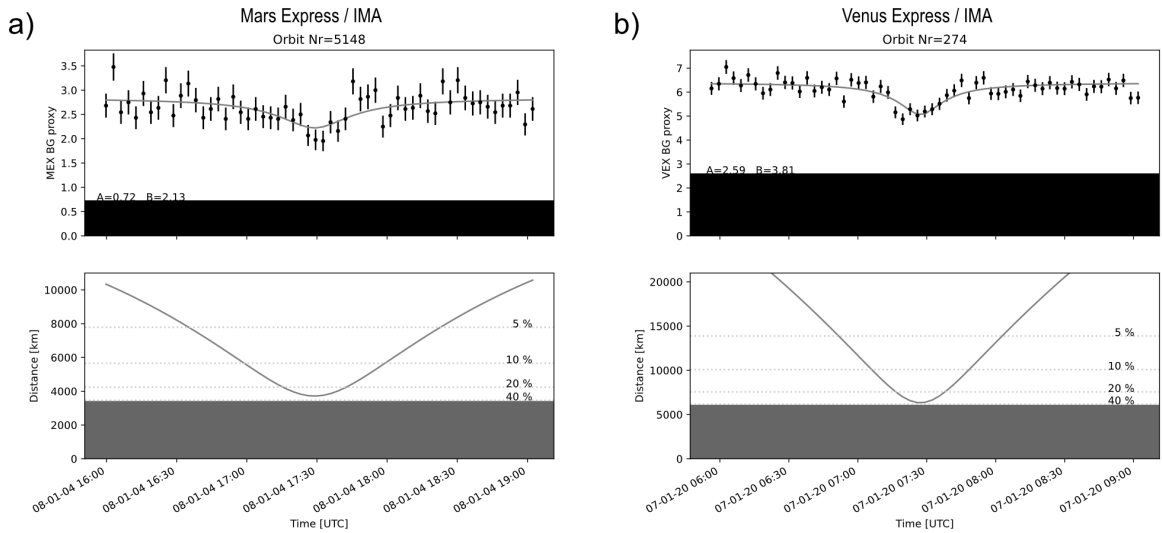

Figure 8: a) Time series of the background counts along an orbit of Mars Express on 2008-01-04 and b) Venus Express on 2007-01-20. (Top) The measured and fitted background counts. The error bar in the top panels represents the counting statistics. The black spans are the β-decay contribution (A). (Bottom) The spacecraft's altitude (from the planetary center). The gray span shows the radius of Venus. The horizontal dotted lines correspond to the altitude with 5%, 10%, 20%, and 40% blockage in our simple geometric shadowing.

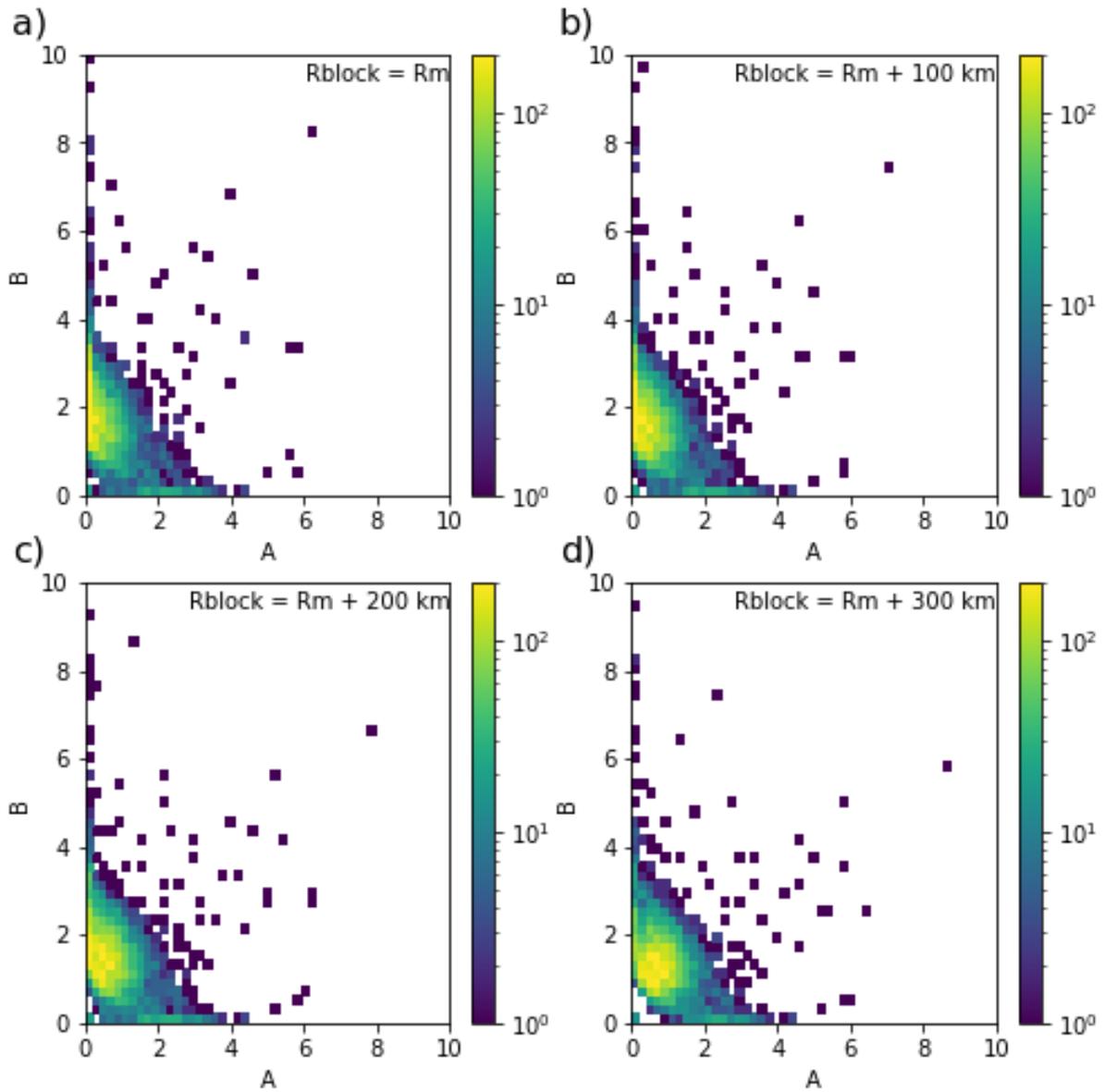

Figure 9: 2-D Histograms of the occurrence of internal (A)–external (B) originated counts measured at Mars. The assumed blockage with the a) radius of Mars ($R_M$=3396 km), b) 3496 km ($R_M$+100 km), c) 3596 km ($R_M$+200 km), and d) 3696 km ($R_M$+300 km).

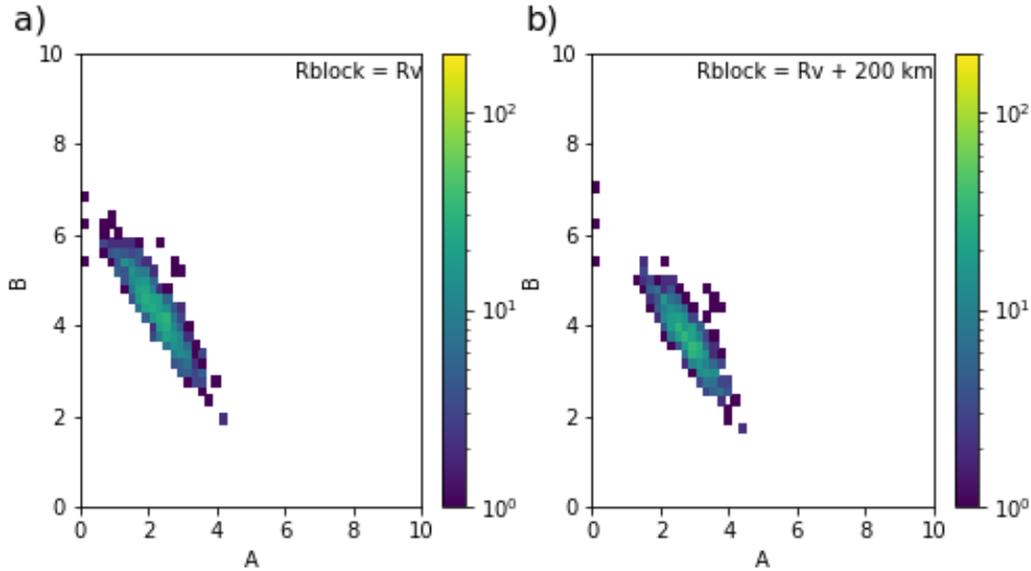

Figure 10: 2-D histogram of calculated external-internal contribution at Venus, in the same format as Figure 9 but with only the obstacle by the sphere with (a) the Venus radius ($R_V$=6051 km) and (b) $R_V$+200 km.

# 6 Concluding remarks

We reported the characteristics of the background counts measured by MCPs as a part of the ion instruments IMA onboard Mars Express and Venus Express. The background counts not only show the sporadic solar energetic particle emission previously reported (e.g., Futaana et al. 2008, Ramstad et al. 2018), but also describe the long-term trends of the GCR intensity. The background datasets at Mars and Venus exhibit the well-known decadal modulation, anti-correlated to the solar activity. In particular, for Mars Express, the background measurements achieved the dataset over 17 years, showing peculiar solar cycle 24 signatures at Mars. The characteristics measured at Mars for cycle 24 are consistent with those at Earth.

We also reported short-term modulations along a spacecraft trajectory caused by changing the blocking effects of the planetary bodies. A simple model of GCR absorption is developed and applied to the data. Clear signatures of planetary blocking are found both for Mars and Venus. Further statistical analysis implies that the effective radius of blocking for a GCR by Mars could be more than 100 km larger than the solid planetary body.

Further investigations using a more sophisticated absorption model are definitely of interest. The spacecraft shape should be considered since a spacecraft also absorbs the GCR flux. For example, a GEANT-4 simulation (Allison et al. 2016) can be used for this purpose. Possibly, the background monitoring might support measuring the residual liquid fuel inside the spacecraft. In addition, we have similar datasets from the electron sensor and energetic neutral atom sensors. These sensors have smaller MCPs than the IMA, so the background counts per unit time are smaller. In addition, there is no obvious "background channel," so careful separation from the real physical foreground counts is needed. However, the advantage is that these sensors have higher time resolution (Posner et al. 2013).

There are not many studies that investigated the MCP interaction with high-energy particles on planetary missions (Andre et al. 2019). The information is critical to exploring the harsh Jupiter environment, such as for the JUICE mission. In addition to robotic missions, high-energy particles are also a thread of human activity in the extra-terrestrial environment. The characterization of high-energy particles using the MCP background from other previous missions and instruments will potentially contribute to the study of the high-energy particle environment by increasing the spatial and temporal coverages. In fact, background counts contain extensive information entropy due to their randomness. Because of a worse compression rate, the background noise requires high telemetry that is a valuable resource for space exploration. In addition, the random background counts increase the load in the internal data processing. This resulted in the decision by the ASPERA-4 team to reduce the background count onboard in 2010. However, we wish to emphasize that the background counts from space instrumentation are not just costly junk. They contain valuable information that is not even the primary scientific objectives of the respective sensors. Throwing away the background data is thus "*mottainai.*"

## Acknowledgment

The Mars Express and Venus Express missions are/were operated by the effort of the European Space Agency (ESA). The ASPERA-3 and ASPERA-4 instrument suites were developed and are/were operated by the Swedish Institute of Space Physics (IRF), with strong international cooperation. The development and operation are/were funded by the Swedish National Space Agency (SNSA). Part of this work was funded by the European Union's Horizon 2020 research and innovation programme under grant agreement No 871149.

The ASPERA data used in this study can be obtained from the FAIR-complied NASA Planetary Data System (PDS) (https://pds-geosciences.wustl.edu/missions/mars_express/aspera.htm). The identical datasets are also available from the Planetary Science Archive (PSA) (https://www.cosmos.esa.int/web/psa/mars-express) and the SPICE kernel repository (https://www.cosmos.esa.int/web/spice/home) at ESA. We acknowledge the NMDB database (www.nmdb.eu), founded under the European Union's FP7 programme (contract no. 213007), and the Thule neutron monitor for providing data, the University of Delaware Department of Physics and Astronomy and the Bartol Research Institute, for providing the excellent data for this study. The sunspot number data were retrieved from the World Data Center SILSO, Royal Observatory of Belgium, Brussels.

This research made use of NumPy (Harris et al. 2020), SciPy (Virtanen et al. 2020), iPython and Jupyter (Perez and Granger 2007), matplotlib (Hunter 2007), and pandas (Mckinney 2011).